\documentclass[%
prl,
twocolumn,
superscriptaddress,
 amsmath,amssymb,
 aps,
]{revtex4-1}

\usepackage{graphicx}
\usepackage{dcolumn}
\usepackage{bm}

\def\ket#1{\mathinner{|{#1}\rangle}}

{\catcode`\|=\active
  \gdef\Braket#1{\begingroup
     \ifx\SavedDoubleVert\relax
       \let\SavedDoubleVert\|\let\|\BraDoubleVert
     \fi
     \mathcode`\|32768\let|\BraVert
     \left<{#1}\right>\endgroup}
}
\def\BraVert{\@ifnextchar|{\|\@gobble}
     {\egroup\,\mid@vertical\,\bgroup}}
\def\BraDoubleVert{\egroup\,\mid@dblvertical\,\bgroup}
\let\SavedDoubleVert\relax

\begin{document}
 \title{Coherence and Raman sideband cooling of a single atom in an optical tweezer}

\author{J.D. Thompson$^{*}$}
\affiliation{Department of Physics, Harvard University, Cambridge,
MA 02138}

\author{T.G. Tiecke$^{*}$}
\affiliation{Department of Physics, Harvard University, Cambridge,
MA 02138}
\affiliation{Department of Physics, MIT-Harvard Center for
Ultracold Atoms, and Research Laboratory of Electronics,
Massachusetts Institute of Technology, Cambridge, Massachusetts
02139}

\author{A.S. Zibrov}
\affiliation{Department of Physics, Harvard University, Cambridge,
MA 02138}

\author{V. Vuleti\'{c}}
\affiliation{Department of Physics, MIT-Harvard Center for
Ultracold Atoms, and Research Laboratory of Electronics,
Massachusetts Institute of Technology, Cambridge, Massachusetts
02139}

\author{M.D.\ Lukin}
\affiliation{Department of Physics, Harvard University, Cambridge,
MA 02138}

 \email{email@place.edu}
 \date{\today}

 \begin{abstract}
We investigate quantum control of a single atom in an optical tweezer trap created by a tightly focused optical beam. We show that longitudinal polarization components in the dipole trap arising from the breakdown of the paraxial approximation give rise to significant internal-state decoherence. We show that this effect can be mitigated by appropriate choice of magnetic bias field, enabling Raman sideband cooling of a single atom close to its three-dimensional ground state in an optical trap with a beam waist as small as $w=900$~nm. We achieve vibrational occupation numbers of $\bar{n}_r = 0.01$ and $\bar{n}_a = 8$ in the radial and axial directions of the trap, corresponding to an rms size of the atomic wavepacket of 24 nm and 270 nm, respectively. This represents a promising starting point for future hybrid quantum systems where atoms are placed in close proximity to surfaces.
\end{abstract}


\maketitle

Single atoms in ``optical tweezer" traps \cite{Schlosser:2001vy} are a promising resource for various applications in quantum science and engineering. They can be individually moved \cite{Beugnon:2007eg}, manipulated \cite{2006PhRvL..96f3001Y,Jones:2007jy}, and read-out \cite{Fuhrmanek:2011vp} in a manner similar to trapped ions. At the same time, they may be strongly coupled to photonic \cite{Alton:2010we, Vetsch:2010tl}, plasmonic \cite{Chang:2009tz}, or other solid-state systems \cite{Hunger:2010fv,Hafezi:2012gz,Rabl:2006vd}, opening a new frontier for the realization of quantum networks and hybrid quantum systems. These intriguing applications require trapping single ultra-cold atoms near surfaces at distances well below an optical wavelength. While this is challenging for ions \cite{Daniilidis:2011dj}, and magnetically trapped atoms \cite{Lin:2004wp,Hunger:2010fv}, it is readily achievable with neutral atoms in optical dipole traps.

The collisional blockade regime \cite{Schlosser:2001vy} of optical dipole traps is an attractive starting point for such experiments because it provides a simple way to load and tightly confine single atoms starting with only a conventional optical molasses. In several experiments, it has been found that the temperature of an atom loaded from a molasses into an optical dipole trap in the collisional blockade regime is in the range 30 $\mu$K to 180 $\mu$K \cite{Tuchendler:2008il, Vetsch:2010tl, 2006PhRvL..96f3001Y, GaEumlTan:2009cu, Jones:2007jy, Rosenfeld:2008iz,Urban:2009vt}. This elevated temperature compared to free-space cooling has been identified as a limiting factor in many recent experiments, as the thermal motion reduces the coherence time \cite{2006PhRvL..96f3001Y,Jones:2007jy,Rosenfeld:2008iz} and impedes full quantum control \cite{Tey:2009we,Urban:2009vt}. Moreover, lower temperatures and a reduction in the spatial extent of the atomic wavepacket are necessary to control the atom at sub-wavelength distances from a surface, or to implement proposed quantum gates using collisional interactions between two ground state atoms \cite{Dorner:2005ed}.

One major challenge to laser cooling and quantum control are polarization effects associated with the breakdown of the paraxial approximation in very tightly focused optical dipole traps. Vector light shifts arising from elliptical light polarization \cite{Corwin:1999ti} are known to be a major obstacle to cooling and manipulating atoms in optical dipole traps. In the paraxial limit, the vector light shift can be eliminated by using a linearly polarized trapping beam. However, in the tightly focused regime, non-paraxial effects produce a longitudinal and spatially inhomogeneous polarization component near the focus that cannot be eliminated. The resulting state-dependent trapping potential leads to dephasing of internal state superpositions \cite{Kuhr:2005tj} and fluctuating dipole force heating, which impairs internal state manipulation as well as laser cooling.

In this Letter, we present a detailed study of the longitudinal polarization component of dipole trap formed by a high numerical aperture lens, demonstrate how the undesirable effects arising from the longitudinal polarization can be partially compensated using a properly oriented magnetic bias field, and apply these results to perform Raman sideband cooling of a single atom. After cooling, the atom is in the ground state along the two radial directions ($\bar{n}_r = 0.01 ^{+0.06}_{-0.01}$), and occupies just a few quantum states ($\bar{n}_a = 8.1(8)$) in the axial trap direction. The corresponding rms size of the atomic wavepacket is given by the ground state length of 24 nm in the radial directions, and a thermal extent of 270 nm in the axial direction, which represents a hundred-fold reduction in spatial volume, and a reduction by $10^4$ in phase-space volume, over the starting conditions.

\begin{figure*}
\begin{center}
\centering
\includegraphics[width=7in]{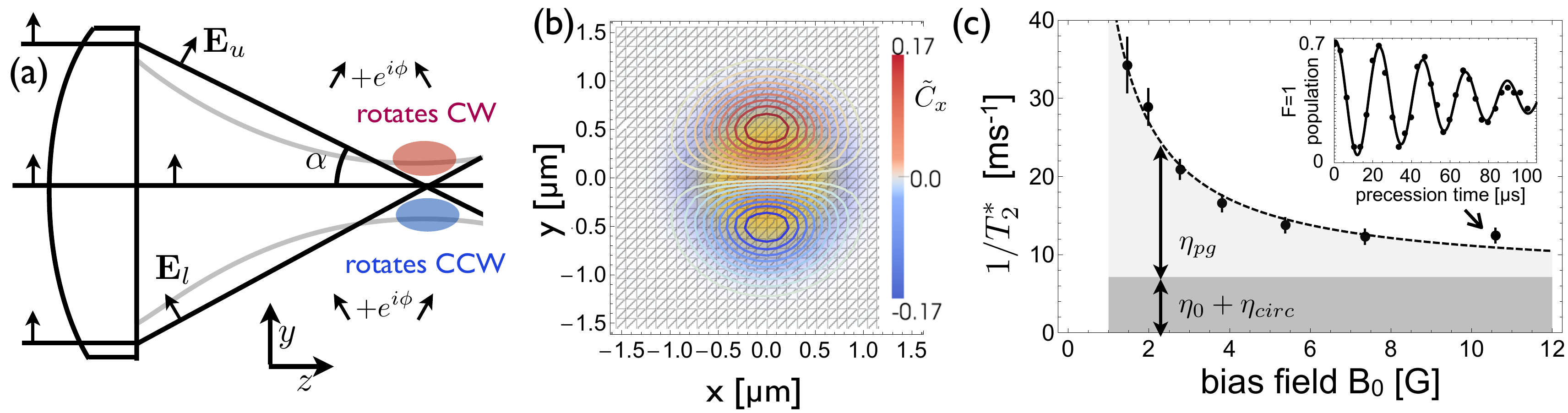}
\caption{(a) Diagram showing the origin of elliptical polarization near the focus (see text). (b) Cut through the focal plane for $\alpha=0.43$. Contour lines show $\tilde{C}_x$, which is $C_x$ scaled to the local intensity $|E(\vec{r})|^2/|E(\vec{r}_{max})|^2$. Shading shows Gaussian intensity profile.  (c) Dephasing rate between the states $\ket{1}$ and $\ket{2}$ as a function of bias field along the Z axis, with a trap wavelength $\lambda=815$ nm. The improvement at large bias fields is due to suppression of the polarization gradient. Fit is to model described in text: $\eta_0 + \eta_{circ}$ are background dephasing rates from the finite detuning and slight elliptical polarization of the dipole trap; $\eta_{pg}$ arises from the longitudinal polarization. Inset: Ramsey measurement of dephasing rate between $\ket{1}$ and $\ket{2}$ at $B_0 = 10.5$ G.}
\label{fig2}
\end{center}
\end{figure*}

The origin and effects of the longitudinal polarization component can be understood in the framework of ray optics (see Figure \ref{fig2}a). The light entering a lens consists of parallel rays with linear polarization transverse to their propagation direction. Upon passing through the lens, rays are refracted according to their distance from the optical axis, and their polarization directions must also deflect to remain transverse to the ray \cite{Richards:1959jw}. In the diffraction-limited volume around the focus, all of these rays interfere and the resulting field is elliptically polarized. Following Fig. \ref{fig2}a, two features of the polarization near the focus emerge: (1) the polarization vector is rotating in the plane set by the incident polarization vector and the optical axis, and (2) the sense of this rotation is opposite above and below the optical axis.

For light that is far-detuned compared to the excited-state hyperfine structure, the vector light shift for alkali atoms in the ground state is \cite{Corwin:1999ti,Deutsch:1998vj}:
\begin{equation}
\label{eq1}
U(\mathbf{r}) = - U_0(\mathbf{r}) \frac{\delta_{2} - \delta_{1}}{\delta_{2} + 2 \delta_{1}} \mathbf{C}(\mathbf{r}) \cdot g_F \mathbf{\hat{F}}
\end{equation}
where $U_0(\mathbf{r})$ is the scalar dipole trap potential, $\delta_{1}$ and $\delta_{2}$ are the detunings from the D1 and D2 lines, respectively, $\mathbf{\epsilon}(\mathbf{r})$ is the local (unit norm) polarization vector, $\mathbf{\hat{F}}$ is the total angular momentum operator and $g_F = \left[F(F+1) - I (I+1) + J(J+1)\right]/F(F+1)$. The term $\mathbf{C} = \mathrm{Im} \left[\mathbf{\epsilon}(\mathbf{r}) \times \mathbf{\epsilon}^*(\mathbf{r})\right]$ is a basis-independent way of expressing the normal vector of the polarization ellipse and the degree of ellipticity (with magnitude $\pm 1$ for circularly polarized light; 0 for linear polarization). Using the vector Debye integral \cite{Richards:1959jw}, we have numerically computed the polarization near the dipole trap focus (Fig. \ref{fig2}b). The most important term is the polarization gradient $dC_{x}/dy$. For a lens with numerical aperture $\alpha$, the maximum gradient, occurring at the beam focus, is well approximated by $3.1 \alpha \sin \alpha / \lambda$ if the lens aperture is uniformly illuminated, and $2.6 \alpha \sin \alpha / \lambda$ if the illumination is a Gaussian beam with a $1/e^2$ diameter equal to the lens aperture diameter.

In the experiments presented here, we use an optical dipole trap for $^{87}$Rb atoms operating at $\lambda_T = 815$ nm with a depth $U_0 = 0.82$ mK. Independent measurements of the depth and radial trap frequency ($\omega_r = 2\pi \times 100$ kHz) allow us to extract a $1/e^2$ beam radius $w=900$~nm and an effective lens aperture $\alpha = 0.43$. For these parameters, we expect $dC_{x}/dy = 0.57/\mu$m. Since the state-dependent potential in Equation (\ref{eq1}) is linear in $\mathbf{\hat{F}}$, it produces the same energy shifts as a magnetic field, and $dC_{x}/dy$ can also be expressed as an effective magnetic field gradient with magnitude $B_x' = 1.4$ G/$\mu$m at the trap center. This gradient gives the trapping potential a significant state-dependent component.

Specifically, in the absence of an externally applied magnetic bias field, atoms in different magnetic sublevels experience trapping potentials that are displaced by $\Delta x = \mu_B \Delta (g_F m_F) B_x' / (m \omega_r^2)$, where $\mu_B \Delta (g_F m_F)$ is the difference in the magnetic moment between the two sublevels. For $\Delta (g_F m_F) = 1/2$, $\Delta x = 11$ nm, which is not negligible compared to the ground state length $\sqrt{\hbar / 2 m \omega} = 24$~nm. While this state-dependent displacement could be useful for Raman cooling or other motional state manipulations along this axis \cite{Forster:2009uca,Li:2012tp}, it also leads to rapid internal-state decoherence on the timescale of the radial trap oscillation period.

To circumvent this problem, we can apply a bias magnetic field in a direction orthogonal to $\hat{x}$. In this case, the effective field gradient is suppressed since $B_{tot} = \sqrt{B_0^2 + (B_x' y)^2} \sim B_0 + (B_x'^2/2 B_0) y^2$, and the gradient results in only a state-dependent change in the strength of the harmonic trap potential. Quantitatively, superpositions of magnetic sublevels that experience different trapping potentials of the form $U_1(\mathbf{r}) = (1+\eta)U_2(\mathbf{r})$ are dephased with a coherence time $T_2^* = 0.97 \times 2 \hbar / (k_B T \eta)$ \cite{Kuhr:2005tj}, where $T$ is the temperature of the atom and $k_B$ is the Boltzmann constant. In the presence of a large orthogonal bias field, the polarization gradient contributes to $\eta$ as $\eta_{pg} = \mu_B \Delta (g_F m_F) B_x'^2 / (3 m \omega^2 B_0)$ (the factor of 1/3 results from averaging over the three trap axes). We can use the dependence on $B_0$ to accurately measure $B'$, and also to suppress the dephasing with large $B_0$.

We measure the decoherence between the states $\ket{1} \equiv \ket{F=1, m_F = -1}$ and $\ket{2} \equiv \ket{F=2, m_F = -2}$ using a sequence that starts with loading a single atom into a tweezer trap with a depth of 1.6 mK at zero bias field, then ramping down the trap depth to 0.82 mK as we ramp up the bias field $B_0$ to the desired value. The atom is optically pumped into $\ket{2}$, the hyperfine transition is driven by a two-photon Raman process in a Doppler-free configuration, and the state detection is accomplished using a push-out beam; these details are described in more detail below. $T_2^*$ is extracted from a Ramsey-type measurement, using a fit to the function introduced in \cite{Kuhr:2005tj}.

At each trap wavelength, we fit $1/T_2^* = 1.03\left(\eta_0 + \eta_{circ} + \eta_{pg}\right)(k_B T / 2 \hbar)$. The only free parameters in this fit are the degree of circular polarization in the incident dipole trap beam due to birefringence ($\eta_{circ}$) and the strength of the effective field gradient $B_x'$. The temperature is determined independently ($T=40 \mu$K for this measurement, see below for technique). $\eta_0$ reflects the different trapping potentials for $F=1$ and $F=2$ atoms due to the finite trap detuning, and is approximately given by the ratio of ground state hyperfine splitting to trap detuning. At a trap wavelength of (802,815) nm, we find $B_x' = (2.4, 1.4) \textrm{ G}/\mu$m, corresponding to gradients $dC_x/dy = (0.46(6), 0.54(3))/\mu$m, in reasonable agreement with our estimate of 0.57 /$\mu$m.

 \begin{figure}[bl]
\begin{center}
\centering
\includegraphics[width=3.375in]{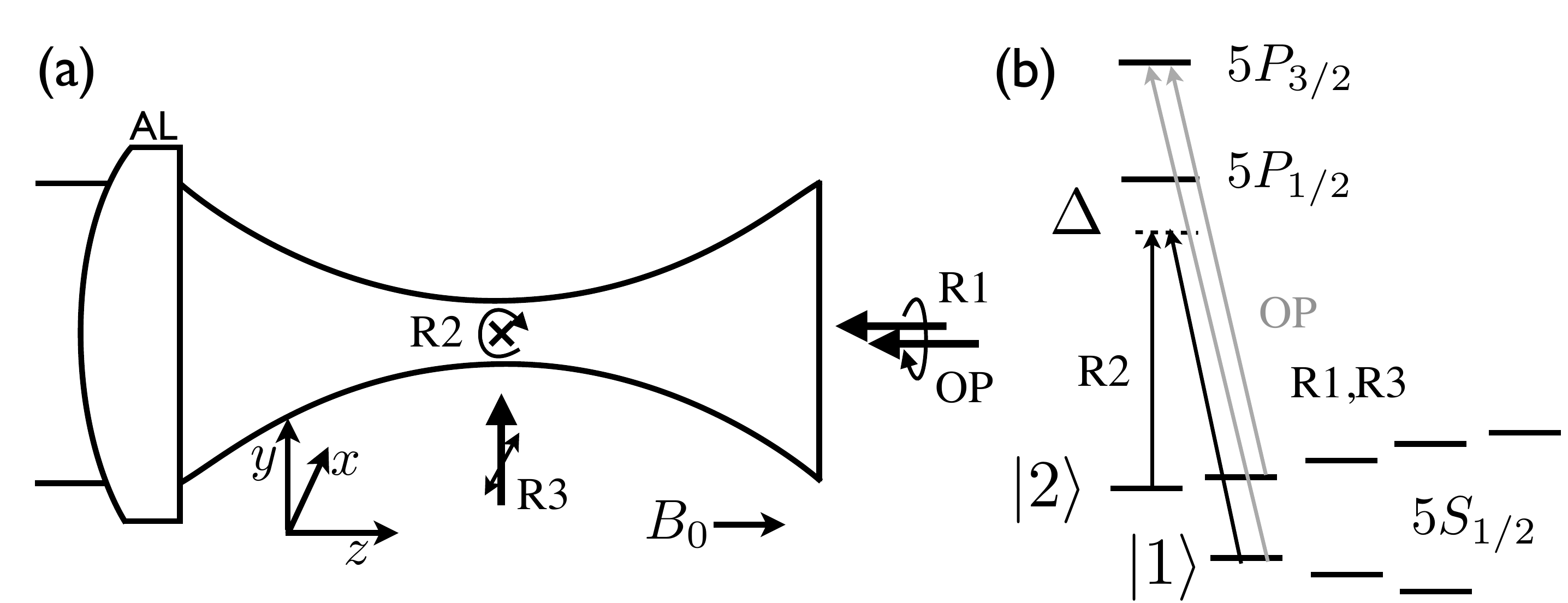}
\caption{(a) Schematic of optical paths. The dipole trap is formed at the focus of an aspheric lens (AL), with a $1/e^2$ radius of 900 nm. Beams R1, R2 (both circularly polarized) and R3 (linearly polarized orthogonally to $\vec{B_0}$) drive Raman transitions. The detuning of the Raman beams $\Delta = 13$ GHz. Optical pumping is along the path marked OP (circularly polarized). (b) States in $^{87}$Rb used for Raman transitions and optical pumping.}
\label{fig1}
\end{center}
\end{figure}

Having developed a detailed understanding of trap-induced decoherence in this system, we now turn to Raman sideband cooling. We use three orthogonal running-wave fields to drive Raman transitions (R1-R3, see Fig. \ref{fig1}a), as in trapped ion experiments \cite{MONROE:1995wt}; the different frequencies are generated by means of an electro-optic modulator \cite{Dotsenko:2004fw}. Optical pumping to the $\ket{2} \def \ket{F=2, m_F=-2}$ state is provided by circularly polarized beams propagating along the magnetic field axis, addressing the $F=1 \rightarrow F'=2$ and $F=2 \rightarrow F'=2$ transitions on the D2 line. The frequencies of the lasers are set to the measured resonances in the dipole trap, which are shifted by $\sim$ 30 MHz from the resonances in free space. The intensities of the two beams about 100 times less than saturation. This ensures that the atom only scatters photons elastically, so heating due to fluctuating dipole forces associated with the (anti-)trapping potential for the excited state is avoided \cite{DALIBARD:1985ug}. For diagnostic purposes, we measure the $F=1$ population by pushing out the atoms in $F=2$ using a beam resonant with the $F=2\rightarrow F'=3$ transition on the D2 line, then measuring whether the atom has remained trapped by turning the molasses back on. This beam is circularly polarized and propagates along the same path as the optical pumping beams.

 \begin{figure}[t]
\begin{center}
\centering
\includegraphics[width=3.375in]{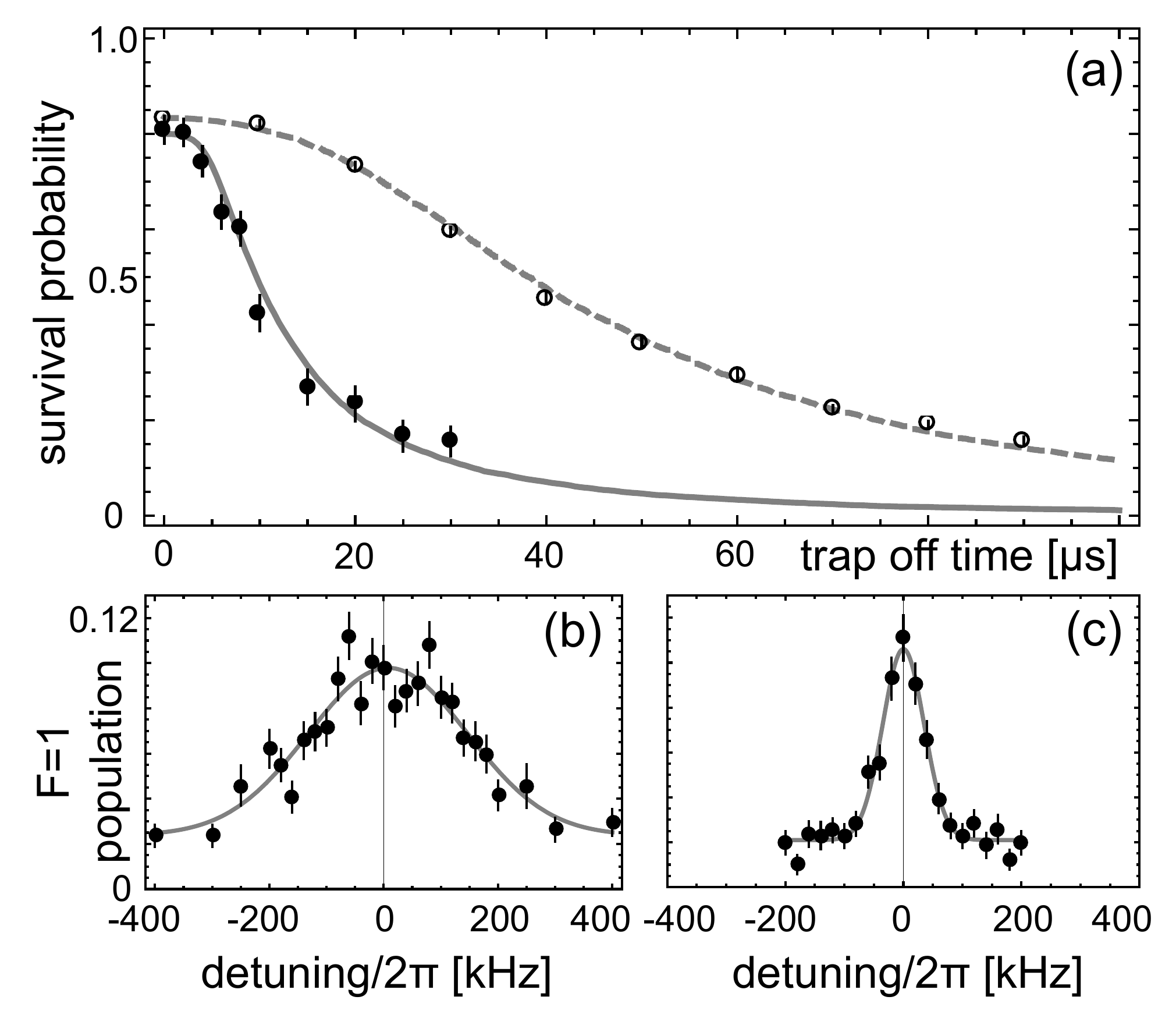}
\caption{(a) Release and recapture temperature measurement. (Closed, open) circles show measurements (before, after) radial cooling. A Monte Carlo model yields kinetic energies $K$ such that $2K/k_B = 52(4) \mu$K before cooling, and $(2K_{r}/k_B,2K_{a}/k_B) = (2.4(1), 158(14)) \mu$K after cooling. (b,c) Measurement of the axial kinetic energy before and after cooling the axial mode, by Doppler spectroscopy. (b) After radial cooling only, $2 K_a/k_B = 129(19) \mu$K. (c) After radial and axial cooling, $2K_a/k_B = 8.1(1) \mu$K.}
\label{fig3}
\end{center}
\end{figure}

In a typical experiment, we load an atom from the MOT into the optical dipole trap with a depth of 1.6 mK at zero bias field, then decrease the trap depth to 0.82 mK while ramping the bias field $B_0$ up to 7.5 G. Lowering the trap depth serves to increase the coherence time while leaving the trap frequencies high enough that sideband cooling is still achievable, with $(\omega_r,\omega_a) = 2 \pi \times (100,15.6)$ kHz. All temperatures reported in this paper are measured in the 0.82 mK deep trap. We cool the atoms in the following sequence: we first apply the R2 and R3 beams (Fig. \ref{fig1}) and the optical pumping beams together for 10 ms to continuously cool the radial modes; then, we perform ten cycles consisting of 2 ms of axial cooling using the R1 and R2 beams, followed by 4 ms of radial cooling using the R2 and R3 beams again. This sequence prevents the radial modes from heating while the axial cooling proceeds.

The parameters for the first radial cooling phase are optimized by measuring the temperature using a release and recapture technique \cite{LETT:1988uk}. This data, shown in Figure \ref{fig3}a, is fit using a Monte-Carlo simulation \cite{Tuchendler:2008il}. The initial kinetic energy $K$ is such that $2K/k_B = 52 \mu$K; the measurement after cooling yields anisotropic kinetic energies of $2K_{r}/k_B = 2.4(1)  \mu$K in the radial direction and $2K_{a}/k_B = 158(14) \mu$K in the axial direction (the release and recapture technique is only weakly sensitive to the axial mode). The fitted kinetic energies represent the global minimum in $\chi^2$ over the entire space of three independent energies for each axis, including unphysical temperatures less than the ground state energy $\hbar \omega / 2 k_B = 2.4 \mu K$ for the radial modes. The agreement of the measured kinetic energy with that of the zero-point motion suggests that we have reached the radial ground state after this cooling phase alone. The radial cooling works best with a two-photon Rabi frequency $\Omega_{R2,R3} = 2\pi \times 17$~kHz and a detuning of $-2\pi \times 100$~kHz from the two-photon resonance, corresponding to  $-\omega_r$, as expected.

\begin{figure}[t]
\begin{center}
\centering
\includegraphics[width=3.375in]{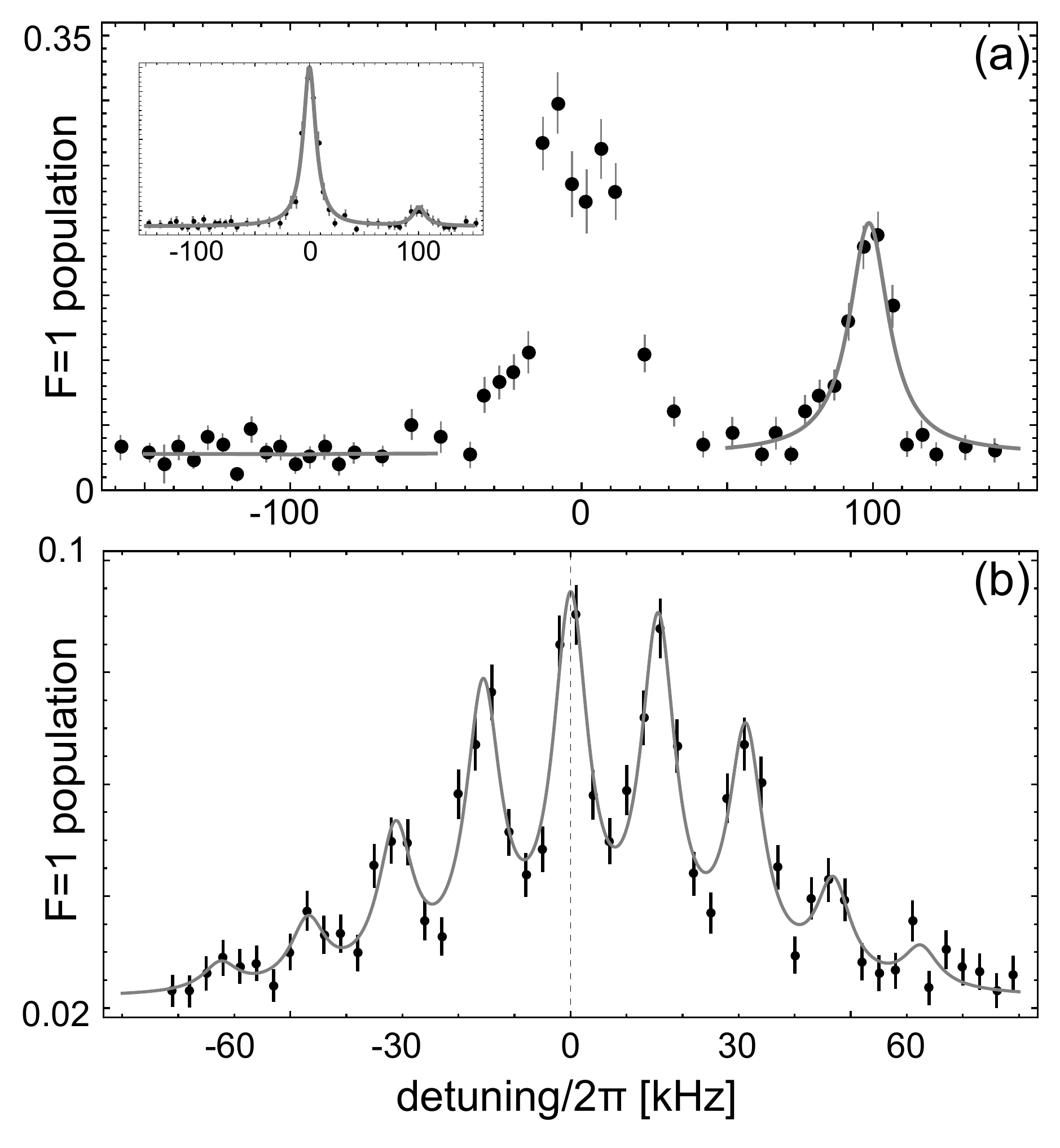}
\caption{Sidebands showing final temperature in the (a) radial and (b) axial directions. In (a), the red and blue sideband amplitudes are fit to independent lorentzians; their ratio yields a radial temperature $\bar{n} = 0.01^{+0.06}_{ -0.01}$. Inset: same measurement with shorter pulse length so the carrier is also resolved. In (b) 9 peaks are fit with independent heights, but equal spacings and widths. The heights are well-described by a thermal distribution with $\bar{n}_a = 8.1(1)$.}
\label{fig4}
\end{center}
\end{figure}

To characterize the axial temperature independently after the radial cooling, we measure the Doppler width of the $\ket{2}$ to $\ket{1}$ transition when driven with the R1 and R2 beams. The wavevector $\Delta \mathbf{k}_{12} = \mathbf{k}_{R1} - \mathbf{k}_{R2}$ has a projection onto the axial and radial directions, but the Doppler profile should mostly be sensitive to the axial mode here since the radial degrees of freedom are already cold. After the first stage of radial cooling, we measure a kinetic energy of $2K_{a}/k_B = 129(19) \mu$K (Fig. \ref{fig3}b). After optimization, we obtain a feature with a width corresponding to $2K_{a}/k_B = 8.1(1) \mu$K (Fig. \ref{fig3}c). This data is fitted to a Gaussian, which conservatively assumes no power broadening. The optimum cooling parameters are a two-photon Rabi frequency of $\Omega_{R1,R2} \sim 2 \pi \times 5$ kHz and a detuning of $-2 \pi \times 60$ kHz. The parameters used for the interleaved radial cooling phases are the same as above.

To obtain more precise measurements of the final temperature of the atom, we resolve the asymmetric motional sidebands along two axes. The ratio of the sideband amplitude gives information about the vibrational state occupation of the atom \cite{MONROE:1995wt}.

Figure \ref{fig4}a shows the sidebands measured in the radial direction with small $\Omega_{R2,R3}$. The blue sideband is essentially absent, with a fitted amplitude 100 times smaller than the red sideband. From this, we extract a final temperature for the radial degrees of freedom of $\bar{n} = 0.01^{+0.06} _{-0.01}$. We do not know to what extent the two radial modes are non-degenerate or what the preferred axes are, but from the release-and-recapture data showing that both modes must be very cold, and the fact that the spectrum shown here does not change if we measure it at a different time after the cooling (up to 100 ms later), we infer that the two modes are not perfectly degenerate and the R2+R3 beams address both modes. Therefore, we conclude that this spectrum reflects the temperature of both radial modes.

We also resolve the axial motional sidebands using the R1 and R2 beams at very low power, and observe a spectrum with nine peaks that is slightly asymmetric (Fig. \ref{fig4}b). We find that the ratios of the measured peak heights correspond very well to a thermal distribution $\rho_{nn} \propto \exp(-n/\bar{n}_a)$ with a mean vibrational number $\bar{n}_a = 8.1(1)$. The corresponding energy $(\bar{n}_a + 1/2) \hbar \omega_a= 6.5$ $\mu$K  $\times k_B$  is similar to the result of the Doppler measurement above.

Several properties of the cooled atom are worth noting. The heating rate for the radial degrees of freedom is very low: we observe heating $\Delta \bar{n} < 0.3$ over 200 ms, which is consistent with what is expected from photon scattering only. We also do not observe radial heating if we translate the atom over distances $\sim 20 \mu$m in $\sim 10$ ms using a scanning galvanometer mirror. Lastly, the fraction of atoms lost from our dipole trap ($\sim$ 30 \%) during the Raman cooling time (150 ms) is due entirely to background gas collisions ($P \sim 10^{-8}$ torr), and not to the cooling process itself. We find that decreasing the Rabi frequency $\Omega_{R1,R2}$ and detuning during the last cooling phase does not decrease the final axial temperature. This is possibly due to one of the following considerations: (1) in our apparatus, the R1+R2 beam pair couples motion along different axes, making it difficult to isolate the axial direction; (2) the ground state Lamb-Dicke factor is close to one on this axis ($k\Delta z = 0.72$); and (3) the optical pumping beams are oriented along the axial direction, potentially causing extra heating which we have not characterized. At the same time, we are not aware of any fundamental effects that would prevent cooling to the ground state in this system.

We have quantified the coherence properties of single atoms subject to the polarization distortions present in tightly-focused and near-field optical dipole traps, and shown that the unwanted effects may be mitigated by applying a magnetic bias field in the appropriate direction. Using this result, we performed Raman sideband cooling of a single atom in an optical tweezer trap close to its ground state, achieving final mean occupation numbers $(n_r, n_r, n_a) = (0.01,0.01,8.1)$.  This technique should find immediate application to a variety of experiments using optical dipole traps.

We acknowledge funding from the NSF, CUA, DARPA, AFOSR, MURI and the Packard Foundation. JDT acknowledges support from the Fannie and John Hertz Foundation and the NSF GRFP.

\emph{Note} After completion of this work, we have become aware of a related demonstration of Raman sideband cooling in an optical tweezer \cite{Kaufman:2012vt}.

\bibliographystyle{apsrev}
\bibliography{ramancoolingv47}

\end{document}